\documentclass[
    11pt,
    final,
    tightenlines,
    notitlepage,
    twoside,
    onecolumn,
    nobibnotes,
    nofootinbib,
    superscriptaddress,
    noshowpacs,
    noshowydk,
    showdoi,
    centertags,
    maik
]{revtex4n}

\usepackage{etoolbox}
\usepackage{xparse}

\usepackage{mathtools}
\usepackage{physics}

\usepackage[inline]{enumitem}

\usepackage{todonotes}

\usepackage{acronym}

\input jpack_e
\setcaptionmargin{5mm}
\onelinecaptionstrue

\columntovsizetrue
\allowdisplaybreaks

\begin{document}

\newtheorem{teo}{Theorem}
\newtheorem*{teon}{Theorem}
\newtheorem{lem}{Lemma}
\newtheorem*{lemn}{Lemma}
\newtheorem{prp}{Proposition}
\newtheorem*{prpn}{Proposition}
\newtheorem{ass}{Assertion}
\newtheorem*{assn}{Assertion}
\newtheorem{assum}{Assumption}
\newtheorem*{assumn}{Assumption}
\newtheorem{stat}{Statement}
\newtheorem*{statn}{Statement}
\newtheorem{cor}{Corollary}
\newtheorem*{corn}{Corollary}
\newtheorem{hyp}{Hypothesis}
\newtheorem*{hypn}{Hypothesis}
\newtheorem{con}{Conjecture}
\newtheorem*{conn}{Conjecture}
\newtheorem{dfn}{Definition}
\newtheorem*{dfnn}{Definition}
\newtheorem{problem}{Problem}
\newtheorem*{problemn}{Problem}
\newtheorem{notat}{Notation}
\newtheorem*{notatn}{Notation}
\newtheorem{quest}{Question}
\newtheorem*{questn}{Question}

\theorembodyfont{\rm}
\newtheorem{rem}{Remark}
\newtheorem*{remn}{Remark}
\newtheorem{exa}{Example}
\newtheorem*{exan}{Example}
\newtheorem{cas}{Case}
\newtheorem*{casn}{Case}
\newtheorem{claim}{Claim}
\newtheorem*{claimn}{Claim}
\newtheorem{com}{Comment}
\newtheorem*{comn}{Comment}

\theoremheaderfont{\it}
\theorembodyfont{\rm}

\newtheorem{proof}{Proof}
\newtheorem*{proofn}{Proof}

\selectlanguage{english}
\Rubrika{\relax}
\CRubrika{\relax}
\SubRubrika{\relax}
\CSubRubrika{\relax}
%

\def\JournalNumber{0}
\def\JournalVolume{00}
%
%
%
\nameVolumeRus{}
\CnameVolumeRus{}
\nameIssueRus{\No}
\CnameIssueRus{}
\namePartRus{}
\namePagesRus{}
\nameYearShortRus{}
\JournalNameRus{}
\TranslitJournalNameRus{}
\JournalName{Regular and Chaotic Dynamics}
\JournalISSNCode{1560-3547}
\IssuePrice{}
\TransYearOfIssue{0000}
\TransCopyrightYear{2016}%
\OrigYearOfIssue{}
\OrigCopyrightYear{2016}%
\OrigIssueNo{\JournalNumber}
\OrigVolumeNo{\JournalVolume}
\TransVolumeNo{\JournalVolume}
\TransIssueNo{\JournalNumber}
\TransPartNo{}
\SHORTjournalPREFIX{RCD} 
\LONGjournalPREFIX{RegDyn} 
\BatFileName{call make_ps.bat} 
\BatSwitch{3} 
\IssueName{}
\SupplementNumber{}
\PublicationSerialNumberInYear{0}
\PublicationSerialNumberInVolume{0}
\ConditionalIssueDate{"year","month","day","name","type"}
\PagePrefix{}
\JournalISSNonlineCode{}
\JournalISSNCodeRus{}
\JournalISSNonlineCodeRus{}
\VolumeName{}
\IssnoName{none}
\PartnoName{}
\FpageNamepp{}
\FpageNnamep{}
\FpagePrefix{}
\LpageNnamepp{}
\LpageNamep{}
\LpagePrefix{}
\VolumePageNumbering{}
\JournalPubID{}
\FirstJournalPageNumber{}
\LastJournalPageNumber{}
\makeatletter
\def\MAIKlogo{RCD Editorial Office}
\def\maikpraefix{10.0000/S}
\edef\@ContentsHeadLineB{Simultaneous English language translation of the journal is available from \noexpand\MAIKlogo}
\def\Distributed{Distributed worldwide by Springer. }
\def\ArticlePages#1{\relax}
\@ifxundefined\CONT@sw{\@booleantrue\CONT@sw}{}%
\@booleantrue\showPACS@sw%
\@booleantrue\showKEYS@sw %
\@booleantrue\noOrigJournalVersion@sw
\@booleantrue\noOrigVolumeNo@sw
\@booleanfalse\noTransVolumeNo@sw
\makeatother
\input maikdoi %

\beginpaper


\input engnames


\definecolor{edits}{RGB}{200,0,0}
\definecolor{strike}{RGB}{130,0,0}
\NewDocumentCommand\EDITS{+m}{{\color{edits}#1}}
\NewDocumentCommand\STRIKE{+m}{{\color{strike}\st{#1}}}


\NewDocumentCommand\eg{}{e.\,g.}
\NewDocumentCommand\ie{}{i.\,e.}
\NewDocumentCommand\cf{}{cf.}

\NewDocumentCommand\MathPeriod{}{\,\text{.}}
\NewDocumentCommand\MathComma{}{\,\text{,}}

\NewDocumentCommand\transpose{}{\mathsf{T}}
\RenewDocumentCommand\vec{m}{\bm{#1}}

\NewDocumentCommand\ETot{}{\ensuremath{\bar{E}_\mathrm{tot}}}
\NewDocumentCommand\Ws{}{\ensuremath{\mathcal{W}_\mathrm{s}}}
\NewDocumentCommand\Wu{}{\ensuremath{\mathcal{W}_\mathrm{u}}}

\DeclarePairedDelimiterX\set[1]{\{}{\}}
    {\NewDocumentCommand\given{}{\;\delimsize\vert\;}#1}


\titlerunning{DYNAMICS AND BIFURCATIONS ON THE NHIM OF A DRIVEN SYSTEM}
\authorrunning{KUCHELMEISTER et al.}
\toctitle{Dynamics and bifurcations on the normally hyperbolic invariant
    manifold of a periodically driven system with rank-1 saddle}
\tocauthor{M. Kuchelmeister}

\title{Dynamics and bifurcations on the normally hyperbolic invariant
    manifold of a periodically driven system with rank-1 saddle}

\articleinenglish
\PublishedInRussianNo

\author{\firstname{Manuel}~\surname{Kuchelmeister}}
\author{\firstname{Johannes}~\surname{Reiff}}
\author{\firstname{J\"org}~\surname{Main}}
\affiliation{
    Institut f\"ur Theoretische Physik I,
    Universit\"at Stuttgart,
    70550 Stuttgart, Germany
}

\author{\firstname{Rigoberto}~\surname{Hernandez}}
\email[Correspondence to: ]{r.hernandez@jhu.edu}
\affiliation{
    Department of Chemistry,
    Johns Hopkins University,
    Baltimore, Maryland 21218, United States
}
\affiliation{
    Departments of Chemical \& Biomolecular Engineering,
    and Materials Science and Engineering,
    Johns Hopkins University,
    Baltimore, Maryland 21218, United States
}

\begin{abstract}
    In chemical reactions, trajectories typically
    turn from reactants to products when crossing a dividing surface
    close to the normally hyperbolic invariant manifold (NHIM) given by
    the intersection of the stable and unstable manifolds of a rank-1
    saddle.  Trajectories started exactly on the NHIM in principle never
    leave this manifold when propagated forward or backward in time.
    This still holds for driven systems when the NHIM itself becomes
    time dependent.  We investigate the dynamics on the NHIM for a
    periodically driven model system with two degrees of freedom by
    numerically stabilizing the motion.  Using Poincar\'e surfaces of
    section we demonstrate the occurrence of structural changes of the
    dynamics, \emph{viz.}, bifurcations of periodic transition state (TS)
    trajectories when changing the amplitude and frequency of the
    external driving.  In particular, periodic TS trajectories with the
    same period as the external driving but significantly different
    parameters---such as mean energy---compared to the
    ordinary TS trajectory can be created in a saddle-node bifurcation.
\end{abstract}

\received{July 12, 2020}
\accepted{September 09, 2020}

\maketitle

\textmakefnmark{0}{)}


\acrodef{BCM}{binary contraction method}
\acrodef{DoF}{degree of freedom}
\acrodefplural{DoF}{degrees of freedom}
\acrodef{DS}{dividing surface}
\acrodef{NHIM}{normally hyperbolic invariant manifold}
\acrodef{PES}{potential energy surface}
\acrodef{PODS}{periodic orbit dividing surface}
\acrodef{PSOS}{Poincar\'e surface of section}
\acrodefplural{PSOS}{Poincar\'e surfaces of section}
\acrodef{TS}{transition state}
\acrodef{TST}{transition state theory}


\section{INTRODUCTION}
\label{sec:intro}

Transition state theory (\acsu{TST})~\cite{
    Pitzer, pech81, truh79, truh85, hynes85b, berne88, nitzan88, rmp90,
    truhlar91, truh96, truh2000, KomatsuzakiBerry01a, pollak05a, Waalkens2008,
    hern08d, Komatsuzaki2010, hern10a, Henkelman2016}
is well established for the computation of rates in systems
with a rank-1 saddle.
In these systems, two different states---\emph{viz.},
reactants and products---are
classified and separated by an appropriately chosen \ac{DS}.
\ac{TST} uses the flux through the \ac{DS} to determine the rate of
a chemical reaction or a similar process.
It has been applied in a broad range of problems in a broad
range of fields including
atomic physics~\cite{Jaffe00},
solid state physics~\cite{Jacucci1984},
cluster formation~\cite{KomatsuzakiBerry99a, KomatsuzakiBerry02},
diffusion dynamics~\cite{toller, voter02b},
cosmology~\cite{Oliveira02},
celestial mechanics~\cite{Jaffe02, Waalkens2005b},
and Bose-Einstein condensates~\cite{
    Huepe1999, Huepe2003, Junginger2012a, Junginger2012b, Junginger2013b}.
For systems, which are time-dependently driven---e.g., by an
oscillating external field---the situation becomes more challenging
because the \ac{DS} itself becomes time dependent and depends
non-trivially on the moving saddle of the potential~\cite{dawn05a, dawn05b}.
The \ac{DS} can nevertheless be obtained
by time-dependent perturbation theory~\cite{hern17f},
through a minimization procedure based on a Lagrangian descriptor~%
\cite{Mancho2010, Mancho2013, hern17h, hern19a},
and other approaches~\cite{hern18g, hern19a}.
Addressing such perturbations is particularly relevant to
the use of chemical control through external fields~\cite{
    pollak99, borondo10,
    Revuelta2015, murgida2015quantum, Keshavamurthy2015control, hern20m}.

In a system with $d$ degrees of freedom, the time-dependent \ac{DS}
embedded in phase space has dimension $2 d - 1$.
It is attached to a $(2 d - 2)$-dimensional \ac{NHIM},
which has the property that every
particle on the \ac{NHIM} will never leave this manifold when
propagated forward or backward in time.
The importance of the geometric structure---with
respect to bifurcations and other objects---of
the \ac{NHIM} on dynamical systems without driving has
been understood for some time~\cite{Wiggins1994} and continues
to receive attention~\cite{jaffe11,komatsuzaki06a,komatsuzaki13b}.
The bifurcation of the \ac{PODS} in two \acp{DoF}
upon varying the total energy
was revealed by Pechukas, Pollak,
and Child~\cite{pech77, pollak78, pollak80}.
Li \emph{et al.}~\cite{Li09} extended their work to three \acp{DoF}
showing that the invariants of motion of the bath modes
can be used to control the bifurcation of the \ac{NHIM}.
In general, bifurcations of the \ac{NHIM} give rise to its breakdown
leading to switching between the dominant reaction coordinate as a function of total
energy of the system~\cite{komatsuzaki11,komatsuzaki15a},
and further leads to the possibility of experimental control~\cite{komatsuzaki15b}.
In this work, we investigate the degree to which such control can play a
role in driven systems.

In periodically driven systems with only one degree of freedom, the
\ac{NHIM} reduces to a point that oscillates
with the same period as the driving potential~\cite{dawn05a, dawn05b,
    hern06d, Kawai2009a, hern14b, hern14f, hern15a, hern16a, hern16h, hern16i}.
This periodic orbit is the one-dimensional \ac{TS} trajectory.
In systems with two or more degrees of freedom, the structure
of the \ac{NHIM} and the dynamics of trajectories on it becomes
non-trivial.
Slowly reacting particles spend a longer time in the vicinity of the
\ac{NHIM} crossing the \ac{DS} closer to the \ac{NHIM}, and therefore
the dynamics on the \ac{NHIM} is of special interest~%
\cite{DynamicalReactionTheory2011}.

In this paper, we resolve the dynamics on the \ac{NHIM} of a
periodically driven two-dimensional system.
We find that the structural changes in the dynamics
are sensitive to variations in
the amplitude and frequency of the external driving.
Using the tools of dynamical systems theory,
we can resolve the stable structures arising from
periodic orbits and visible as tori in the \ac{PSOS}.
With increasing perturbation through external driving,
we find bifurcations that signal emergent behavior
inside of the \ac{NHIM}.


\section{THEORY AND METHODS}
\label{sec:theory}

A pair of fundamental assumptions in the original construction of \ac{TST}
is that chemical reaction kinetics can be modeled
by nuclear motion on a Born-Oppenheimer surface,
and that said motion can be described by classical mechanics.
The system can be represented as a multidimensional effective particle.
Its evolution is a trajectory on the potential energy surface
connecting reactant and product regions or valleys.
In a typical scenario, there exists a rank-1 saddle between these regions.
The third of \ac{TST}'s assumption is that there exists a
\ac{DS} associated somehow with this \ac{TS}\footnote{
    Note that we use the phrase \emph{transition state} to refer to
    the ensemble of states bound indefinitely to the saddle region, \ie,
    those located on the codimension-2 \ac{NHIM}.
    This is distinct from the codimension-1 \emph{dividing surface}.
    See footnote~6 of Ref.~\cite{Waalkens2008} for a more detailed discussion.}
whose associated directional flux weighted
by the (reactant) population is a rate constant.
This connection is exact only if the \ac{DS} is recrossing-free.
Below, we review the geometric approaches that have been
developed to construct it.
An additional layer of complexity arises when the system is driven
by time-dependent potentials~\cite{hern08d}.
A recent method for the construction of the associated time-dependent
\ac{DS} in 1-\ac{DoF} is also discussed below.
Together, this section sets the stage for the elaboration of these methods
in the context of the 2-\ac{DoF} system that form the central results presented
in the next section.


\subsection{Model system}
\label{sec:results/model}

\begin{figure}[tbp]
    \centering
    \includegraphics[width=\linewidth]{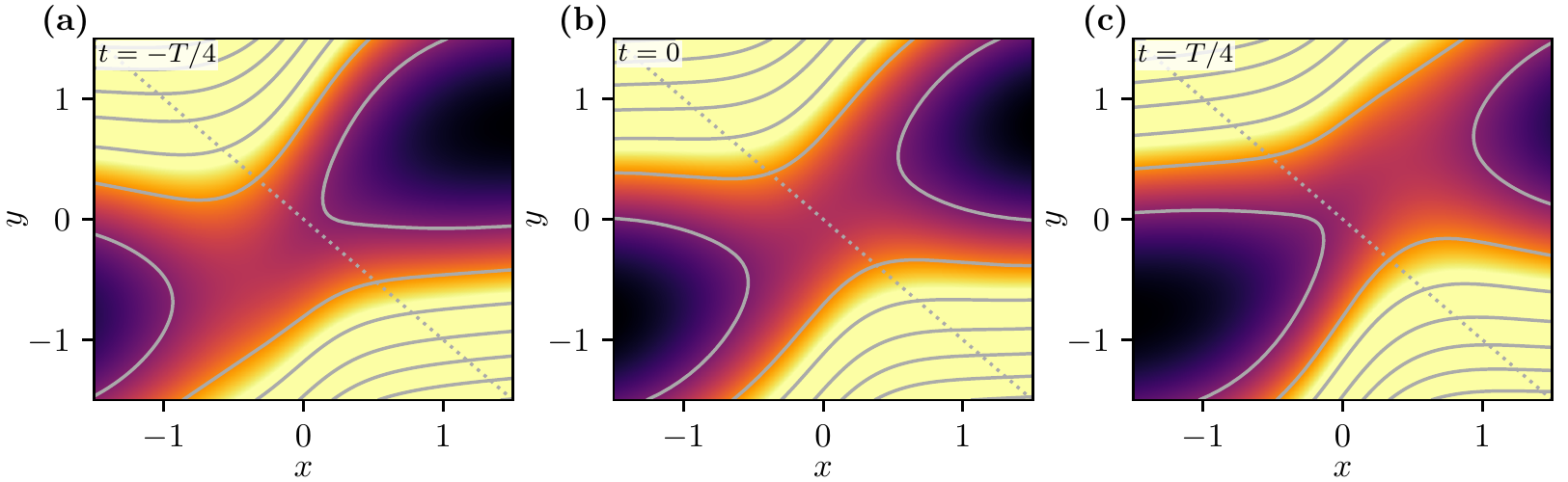}
    \caption{%
        Time-dependent potential $V(x, y, t)$ of the model
        described by Eq.~\eqref{eq:results/potential}
        with $\hat{x} = 0.4$
        and $\omega_x = 2 \pi / T$
        at three different times $t$ given in the figures.
        The dotted diagonal line serves as a guide to the eye.}
    \label{fig:potential}
\end{figure}

For the investigation of the dynamics on the \ac{NHIM}
and the dependence of the dynamics on parameters of the external driving,
we use a two-dimensional model system
which has already been studied in
previous publications~\cite{hern19a, hern19e, hern20d}.
The time-dependent potential
\begin{equation}
    \label{eq:results/potential}
    V(x, y, t)
    = 2 \exp{-[x - \hat{x} \sin(\omega_x t)]^2}
        + 2 \qty[y - \frac{2}{\pi} \arctan(2 x)]^2
\end{equation}
describes a two-dimensional, periodically oscillating energy landscape
including a rank-1 saddle.
The saddle's unstable direction is constructed
via a Gaussian barrier along the approximate \emph{reaction coordinate} $x$.
To make the saddle time-dependent,
the barrier's position is driven time-periodically.
Both the driving amplitude $\hat{x}$ and frequency $\omega_x$
will be varied in the following sections.
In order to expand the system to two \acp{DoF},
a harmonic oscillator is added through an \emph{orthogonal mode} $y$.
This new \ac{DoF} is coupled via the nonlinear term $\arctan(2 x)$,
resulting in non-separable dynamics in the vicinity of the reaction
without changing the properties on the reactant and product regions.
Further system parameters,
such as the strength of the harmonic oscillator or the saddle height and width,
are not changed throughout this paper and are therefore set to fixed values.
This prototypical potential,
which can be used to describe chemical reactions qualitatively,
is depicted in Fig.~\ref{fig:potential} at three different points in time $t$.
In the following we investigate the dynamics on the \ac{NHIM}
parameterized by the orthogonal modes $y$, $v_y$, and time $t$.


\subsection{Basics of transition state theory}
\label{sec:theory/tst}

\begin{figure}[tbp]
    \centering
    \includegraphics[width=\linewidth]{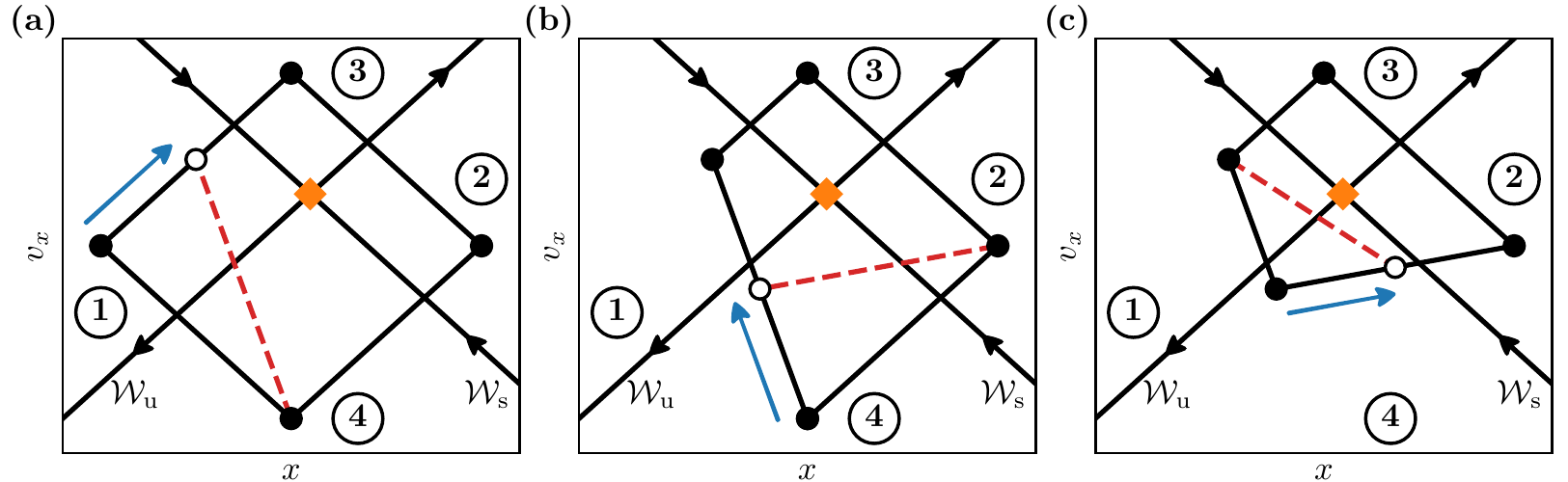}
    \caption{%
        Use of the binary contraction method to uncover the \acs{NHIM}
        in phase space.
        The stable \Ws\ and unstable \Wu\ manifolds associated with the barrier
        divide the phase space into distinct regions labeled 1 through 4.
        The \acs{NHIM} is located at the intersection
        of the two manifolds' closures
        and is marked with a diamond.
        To numerically obtain this point,
        a quadrangle with one corner in each region is set up.
        In turn, the midpoint for every edge is determined,
        shown as an empty circle.
        Depending on which region the midpoint is in,
        it subsequently replaces one of the previous corners.
        This procedure is repeated
        until the quadrangle is sufficiently small.
        Frames a, b, and c show the first three steps of the method.}
    \label{fig:reactive_regions}
\end{figure}

We begin by considering
the time-invariant case of a reactive
1-\ac{DoF} system
with an energy barrier along the reaction coordinate $x$
separating reactants from products.
Depending on the initial conditions and therefore the energy,
most particles can be classified as either
\begin{enumerate*}[label=\arabic*)]
    \item non-reactive reactants,
    \item non-reactive products,
    \item reactive reactants, or
    \item reactive products.
\end{enumerate*}
This results in four distinct regions in phase space,
as indicated in Fig.~\ref{fig:reactive_regions}.
Trajectories started at initial conditions within any of these regions
eventually leave the barrier's vicinity
both forward and backward in time.

The regions are separated by two kinds of critical trajectories.
The points
initiating trajectories that approach the barrier top
without reaching it as $t \to \infty$
and leave its vicinity when propagating backward in time
form the stable manifold \Ws;
the reverse is called the unstable manifold \Wu.
Both manifolds belong to a hyperbolic fixed point
at their closures' intersection
called the \ac{NHIM}.
In the special
case of a time-invariant 1-\ac{DoF} Hamiltonian,
the \ac{NHIM} is associated with the unstable trajectory
for
which a particle remains precariously fixed at the barrier maximum.
Such a trajectory is an example of the \ac{TS} trajectory.
We can extend this structure to periodically driven systems.
Now, the \ac{NHIM}---or \ac{TS} trajectory---will
be time-dependent as well,
but it detaches from the barrier's maximum.
Its defining property%
---a set of unstable trajectories
trapped indefinitely in the vicinity of the barrier---%
still applies.

To numerically resolve the reactive and nonreactive regions presented above,
one has to determine in which state---reactant or product---particles
start from, and in which state they end.
This can be defined through the use of an interaction region---namely
an interval along the reaction coordinate $x$---outside of which the system
is \emph{clearly} in the reactant or product regions.
One can then associate initial conditions to one of the regions
by propagating the corresponding particle both forward and backward in time
until it leaves the interaction region to the reactant or product side.

This classification can be used to precisely calculate the position of
the stable and unstable manifolds as well as the \ac{NHIM}.
For the former, a bisection algorithm
which finds the border between two adjacent regions
can be employed.
Since these borders meet
at a hyperbolic fixed point,
the \ac{NHIM}'s position can be determined by linking the four bisections
belonging to the stable and unstable manifolds.
The resulting algorithm is called the \acf{BCM}~\cite{hern18g}
and is illustrated in Fig.~\ref{fig:reactive_regions}.
This conjectured algorithm has been seen to result
in a \ac{NHIM} in all of the chemically-motivated
applications we have employed it.
This is however not guaranteed to converge in diabolical cases
characterized by multiple saddles in proximity, although
it did work if they were sufficiently far apart~\cite{hern20T2}.

The simple picture presented above can be extended
to systems with arbitrarily many \acp{DoF}.
The potential energy landscape in a $d$-\ac{DoF} system
typically features a rank-1 saddle separating reactants from products.
While there is still a single unstable reaction coordinate $x$,
we now have $d - 1$ additional stable orthogonal modes $\vec{y}$.
As a consequence, the stable and unstable manifolds become
hypersurfaces with codimension 1 in phase space.
Similarly, the \ac{NHIM} becomes a codimension 2 manifold
located at the intersection of the closures of \Ws\ and \Wu.
Likewise, the \ac{TS} describes
the ensemble of all trajectories
confined to the \ac{NHIM}.
If coordinates are chosen well,
every $x$--$v_x$ section of the phase space
for a given position $\vec{y}$ and velocity $\vec{v}_y$
shows the cross-like structure shown in Fig.~\ref{fig:reactive_regions}.
The numerical algorithms presented above
can therefore be used without modification
to calculate individual points of the \ac{NHIM}.


\subsection{Dynamics on the NHIM}
\label{sec:theory/dynamics}

The aim of this paper is to uncover and resolve
non-trivial structures and dynamics on the \ac{NHIM}
as revealed below.
Unfortunately, in the case of a 1-\ac{DoF} system,
the dynamics on the \ac{NHIM} is trivial since it only consists of one point.
This single point does not have an internal structure.
If the system is time-invariant,
even a 2-\ac{DoF} system exhibits integrable dynamics on the \ac{NHIM}
because the effective dynamics is one dimensional and the energy is conserved.
In the following, we therefore investigate a time-dependent 2-\ac{DoF} system
as it is large enough to exhibit nontrivial behavior,
but small enough to be visualized.

Despite the fact that the \ac{NHIM} is a mathematically invariant subspace,
trajectories started on it deviate exponentially fast as time goes by
because of limited numerical precision.
Therefore, numerically calculated
trajectories have to be stabilized explicitly on the \ac{NHIM},
\eg\ using the \ac{BCM}.
Therein, after each time step $\Delta t$,
the coordinates $x$ and $v_x$
are recalculated to project
the particle back onto the \ac{NHIM}.
A similar projection has been used in Ref.~\cite{hern20d}.
Care has to be taken
to ensure that this projection does not introduce significant errors.
To achieve this, the projected distance in $x$ and $v_x$ is measured.
If a maximum distance is exceeded, the projection is rejected
and the particle is reset to an earlier time and position.
The propagation is then repeated with a smaller value of $\Delta t$.
In the present case, an empirical maximum distance of
$\sqrt{(\Delta x)^2 + (\Delta v_x)^2} \le 10^{-3}$ has been used.
We found numerically that
smaller thresholds do not change the results in a significant way,
but we have yet to prove a rigorous bound on this threshold.
This procedure effectively reduces the dimensionality on the
subspace traversed by the trajectories from the four dimensions of the
full phase space to the two dimensions of the \ac{NHIM}.
In higher dimensions,
one could instead take advantage of an artificial neural net
as recently done in Ref.~\cite{hern20d}
to stabilize the trajectories at the
price of some loss in accuracy.

\subsubsection{Poincar\'e surfaces of section}
\label{sec:theory/dynamics/psos}

Although the \ac{NHIM} in the present case is only a two-dimensional subspace,
visualizing many trajectories on it can still get confusing quickly.
A reliable method for such a task is the Poincar\'e map,
also known as the \acp{PSOS}~\cite{Lichtenberg82}.
Using this method, the dimensionality of the phase space is reduced
by only showing intersection points with a given sectional surface
instead of the whole trajectory.
When the system has a natural period, such as in the present
case of a periodically driven system,
one can effect a similar dimensional reduction using
a stroboscopic map capturing periodic sections at corresponding
intervals in time.
For a time-dependent, effectively two-dimensional
system,
the stroboscopic \ac{PSOS} corresponds to a set $\Sigma$ given by
\begin{equation}
    \Sigma
    = \set*{\,
        \vec{\gamma}(t_n) \in \mathbb{R}^2
        \given t_n = t_0 + n T, n \in \mathbb{N}_0
    \,}
    \MathComma
\end{equation}
where $\vec{\gamma}(t) = (y(t), v_y(t))^\transpose$
denotes the phase space vector of
the system constrained to the \ac{NHIM}
and $T$ is the period of the driving.
The previously continuous trajectory thereby
becomes discretized under the stroboscopic map.
In the numerical results presented below,
the fixed points of the map are found using the modified Powell method,
\texttt{scipy.optimize.root(method='hybr')},
implemented in the Python package SciPy~\cite{SciPy2020}.

\subsubsection{Regular and chaotic dynamics}
\label{sec:theory/dynamics/reg_chaos}

Dynamical properties, such as the integrability of the system,
can be determined with the stroboscopic map introduced in the previous
section.
Near integrability is revealed
by the existence of torus-like structures in the system's \ac{PSOS}.
Changes in the system's parameters,
can lead to a transition from near integrable to chaotic
as revealed by the emergence of stochastic structure in the
\ac{PSOS}.

A periodic trajectory with the property
\begin{equation}
    \vec{\gamma}_n = \vec{\gamma}_{n + 1}
    \qq{where}
    \vec{\gamma}_n \equiv \vec{\gamma}(t_0 + n T)
    \qand
    n \in \mathbb{N}_0
\end{equation}
manifests as a fixed point with period $T$ in the \ac{PSOS}.
Fixed points with periodicity $s T$, where $s \in \mathbb{N}$,
analogously fulfill $\vec{\gamma}_n = \vec{\gamma}_{n + s}$.
They appear as $s$-cycles in the \ac{PSOS}
since the stroboscopic map records a point once per system period $T$.
The difference $\vec{\gamma}_{n + 1} - \vec{\gamma}_n$ is a vector
whose length increases monotonically with distance from the fixed point
within the fixed point's neighborhood.
This can be exploited to find fixed points
through a root search algorithm.

There are two kinds of fixed points that must be distinguished:
Elliptic fixed points belong to stable periodic orbits.
Trajectories that start in the vicinity of such an orbit
stay in its vicinity indefinitely.
In the \ac{PSOS}, these show up as
concentric, possibly deformed torus-like structures.
Hyperbolic fixed points, on the other hand, correspond to unstable orbits.
Trajectories started in their vicinity act
as if they were being repelled by the fixed point,
leading to hyperbola-shaped structures in the \ac{PSOS}.

When varying a system's parameters,
its dynamics can change qualitatively in so-called bifurcations.
Local bifurcations are identifiable
via the creation or annihilation of fixed points.
Most important for this paper are saddle-node bifurcations
in which a pair of an elliptic and a hyperbolic fixed point emerge or vanish.
A typical path from regular to chaotic, non-integrable dynamics is
via an infinite series of bifurcations.


\section{RESULTS AND DISCUSSION}
\label{sec:results}


\subsection{Dependence of the dynamics on the external driving parameters}
\label{sec:results/driving}

\begin{figure}[tbp]
    \includegraphics[width=\linewidth]{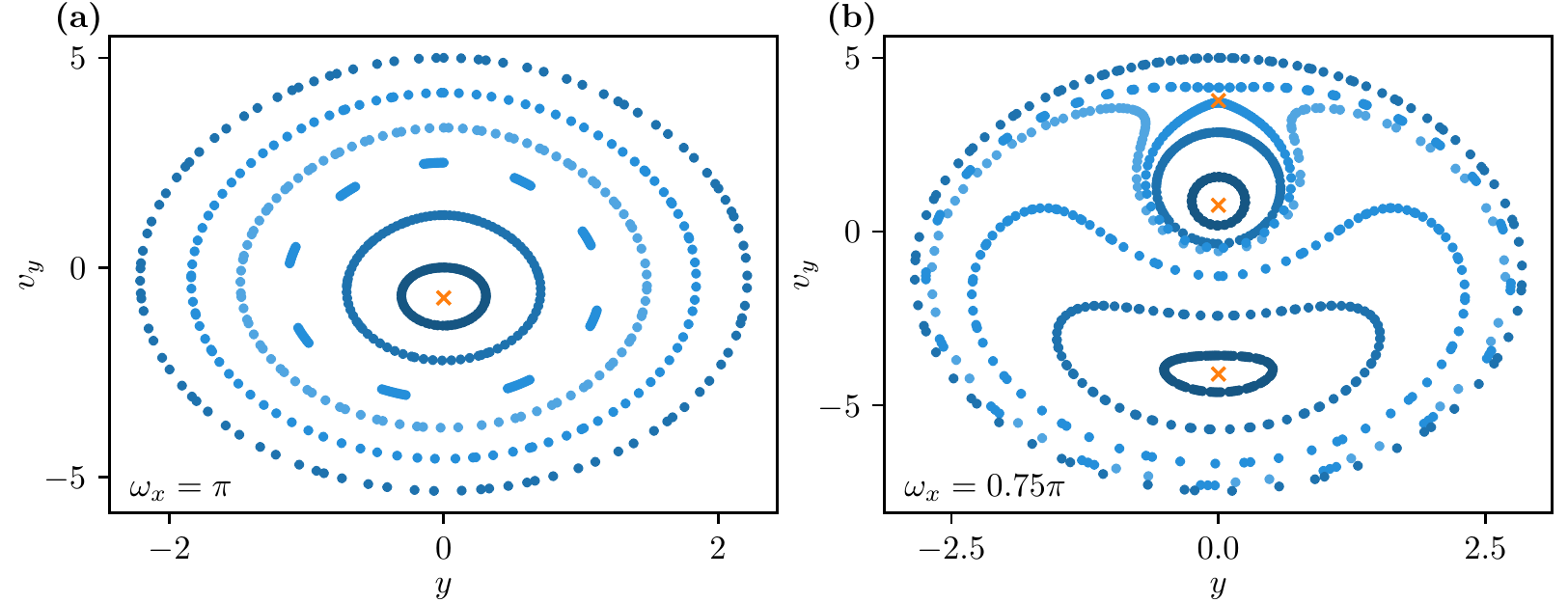}
    \caption{%
        Stroboscopic map of selected trajectories (\acs{PSOS})
        started at time $t_0 = 0$ for different frequencies $\omega_x$.
        The driving amplitude is fixed at $\hat{x} = 0.4$.
        Crosses indicate fixed points.
        A color gradient is used to help distinguish neighboring trajectories.
        (a)~With $\omega_x = \pi$, a single elliptic fixed point
        at $(y = 0, v_y \approx -0.72)$ exists.
        (b)~When decreasing $\omega_x$ to $0.75 \pi$,
        this fixed point moves down to $v_y \approx -4.10$,
        while a pair of elliptic and hyperbolic fixed points
        emerge from a saddle-node bifurcation
        around $\omega_x \approx 0.802 \pi$.}
    \label{fig:psos_bifurcation}
\end{figure}

We start with the driving parameters used in our recent work~\cite{hern19a},
$\omega_x = \pi$ and $\hat{x} = 0.4$.
To construct the \ac{PSOS},
a number of trajectories are started on a specified domain
at $t_0 = 0$
and are propagated for $100$ periods each.
It turns out that a reasonable choice for this domain
is the set of points on the $v_y$ axis at $y = 0$ as that
is enough to map the full subspace of the \ac{NHIM}.
That is, given $y$ and $v_y$, there exits only one
$x$ and $v_x$ which keep the phase-space point on the \ac{NHIM}.
Thus, the trajectories are fully specified for a given
$y$, $v_y$, and $t$
and the restriction of $y = 0$ does
not lose any generality in the \acp{PSOS}
shown in Fig.~\ref{fig:psos_bifurcation}a.

In the stroboscopic map,
we can see a simple structure with a single elliptic fixed point.
The elliptical patterns around it
indicate that the system's dynamics on the \ac{NHIM}
is regular and nearly integrable.
Therefore, trajectories in phase space lie on two-dimensional tori,
for which one approximate constant of motion exists~\cite{Wimberger2014}.
In the neighborhood of the elliptic fixed point,
resonance gaps can be seen,
which indicates the existence of periodic orbits.
Remaining trajectories are quasi-periodic,
meaning they have an irrational winding number
and that for arbitrary long integration times
they would cover the entire torus surface area.

A qualitative change in the system's dynamics occurs in the \ac{PSOS}
when the frequency of driving is decreased to $\omega_x = 0.75 \pi$,
as can be seen in Fig.~\ref{fig:psos_bifurcation}b:
The elliptic fixed point from Fig.~\ref{fig:psos_bifurcation}a
moves down in $v_y$.
In addition, a pair of fixed
points---one elliptic and one hyperbolic---emerge from the saddle-node bifurcation
as shown in Sec.~\ref{sec:results/driving/bifurcations}.
As a result, the elliptic structure of the original fixed point
gets deformed significantly.
While this changes its appearance,
it does not lead to chaotic behavior.

\subsubsection{Saddle-node bifurcations by variation of parameters}
\label{sec:results/driving/bifurcations}

\begin{figure}[tbp]
    \includegraphics[width=\linewidth]{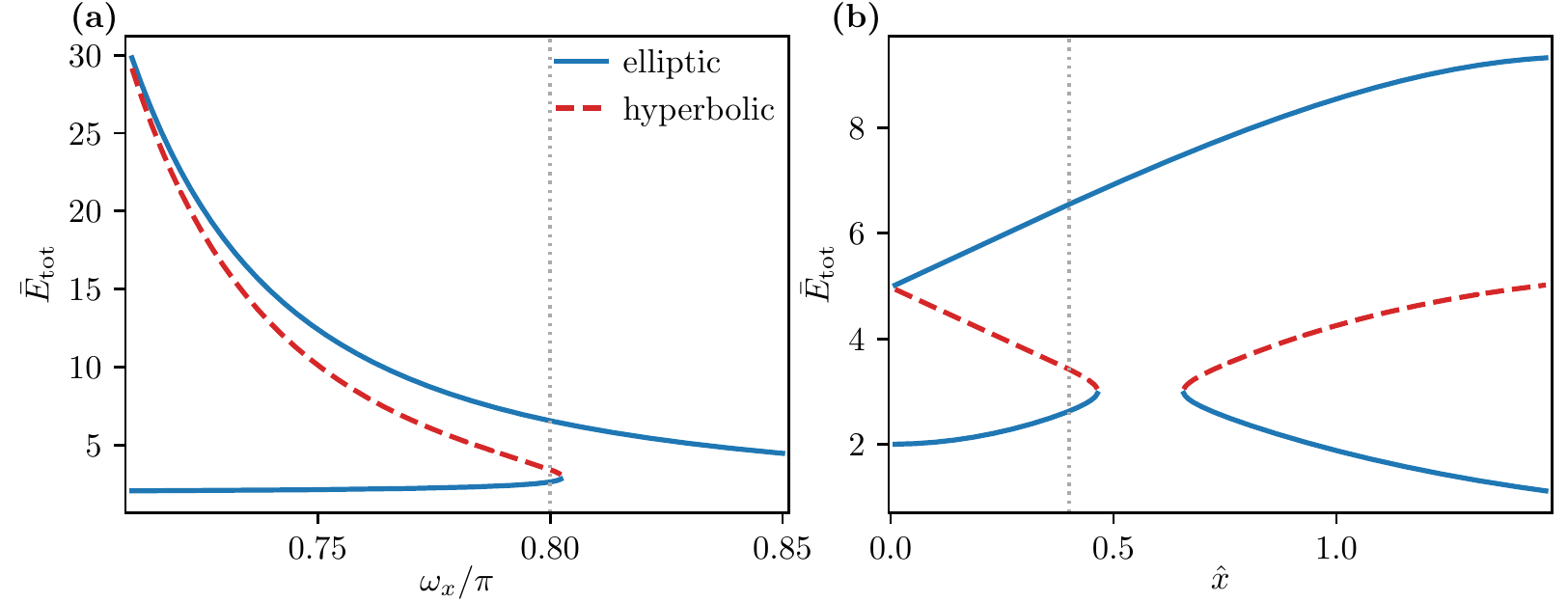}
    \caption{%
        (a)~Average total energy \ETot\ of periodic trajectories
        associated with fixed points (\cf\ Fig.~\ref{fig:psos_bifurcation})
        as a function of driving frequency $\omega_x$.
        The driving amplitude is fixed at $\hat{x} = 0.4$.
        Around $\omega_x \approx 0.802 \pi$,
        a pair of elliptic and hyperbolic fixed points
        annihilate in a saddle-node bifurcation.
        (b)~Average total energy \ETot\ analogous to (a)
        as a function of driving amplitude $\hat{x}$
        for $\omega_x = 0.8 \pi$.
        The system exhibits two saddle-node bifurcations
        around $\hat{x} \approx 0.47$ and $\hat{x} \approx 0.65$.
        At $\hat{x} = 0$, two orbits have
        identical potential energies and reversed velocities.
        Their total energies are therefore degenerate.
        Vertical lines indicate where the parameters of (a) and (b) intersect.}
    \label{fig:energy_bifurcations}
\end{figure}

From Fig.~\ref{fig:psos_bifurcation}, it seems natural to assume that
there is a bifurcation in the parameter range $0.75 \pi < \omega_x < \pi$.
To confirm this assumption,
we need to examine the fixed points systematically.
To do so, fixed points were tracked using a root search algorithm
as described in Sec.~\ref{sec:theory/dynamics/reg_chaos}.

Figure~\ref{fig:energy_bifurcations} displays bifurcation diagrams,
where the averaged total energy \ETot\
of trajectories associated with fixed points
is shown as a function of saddle frequency $\omega_x$
and driving amplitude $\hat{x}$.
In Fig.~\ref{fig:energy_bifurcations}a, the driving frequency is varied between
$0.71 \pi \leq \omega_x \leq 0.85 \pi$.
One can clearly see the annihilation of a pair of fixed points%
---one elliptic and one hyperbolic---%
in a saddle-node bifurcation at $\omega_x \approx 0.802 \pi$.
For decreasing frequencies, the original elliptic fixed point
from Fig.~\ref{fig:psos_bifurcation}a
gets pushed in the negative $v_y$ direction away from the origin,
while the new elliptic fixed point formed in the bifurcation
slowly converges towards the origin.
By contrast, the hyperbolic fixed point's position $v_y$
increases for decreasing driving frequencies.
Thus distinguishing
the elliptic fixed point near the origin for slow and high driving frequencies
must be done with some care.

Bifurcations do not only occur when changing the saddle's frequency,
but also when altering its driving amplitude.
This can be seen in Fig.~\ref{fig:energy_bifurcations}b.
The saddle frequency was set to $\omega_x = 0.8 \pi$,
near the parameter where the bifurcation occurs
in Fig.~\ref{fig:energy_bifurcations}a.
Changing the saddle amplitude from $0 < \hat{x} \leq 1.47$
results in two saddle-node bifurcations,
one annihilating a fixed point pair at $\hat{x} \approx 0.47$
and one creating a fixed point pair at $\hat{x} \approx 0.65$.

\subsubsection{Trajectories in position space for different driving frequencies}
\label{sec:results/driving/trajectories}

The next step in the classification of the dynamics requires
the characterization of the periodic trajectories
associated with the fixed points
found in Sec.~\ref{sec:results/driving/bifurcations}.

\begin{figure}[tbp]
    \includegraphics[width=\linewidth]{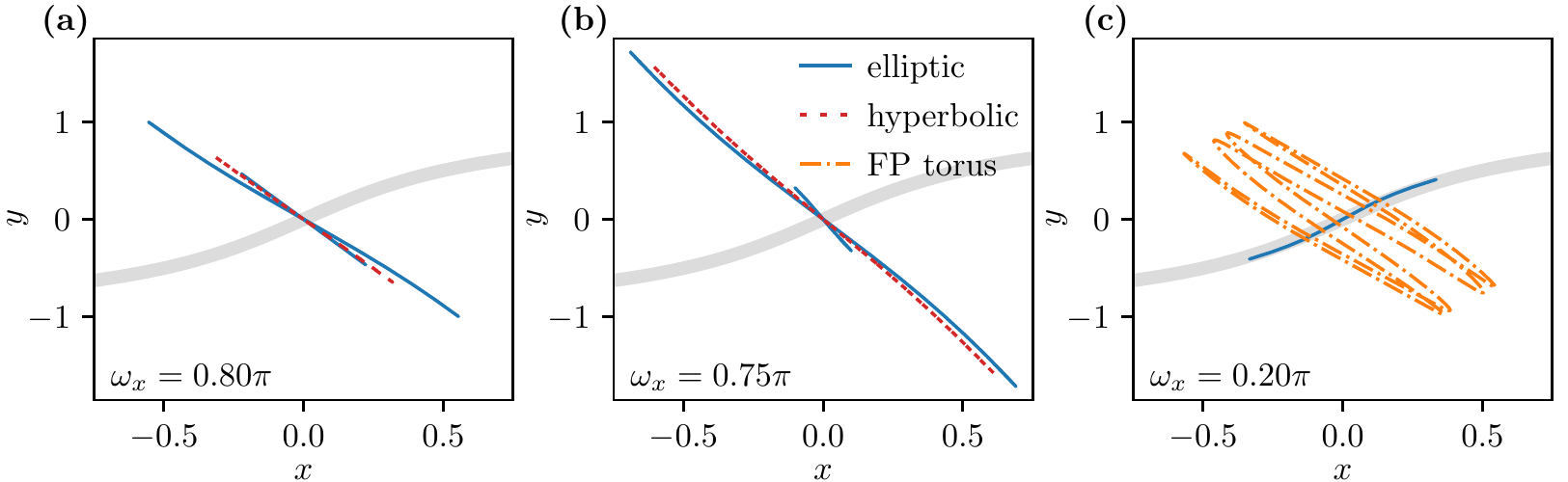}
    \caption{%
        Position $(x, y)$ of periodic orbits on the \acs{NHIM}
        for different frequencies $\omega_x$.
        The driving amplitude is fixed at $\hat{x} = 0.4$.
        The minimum energy path is shown as a thick, light gray line.
        (a)~With a frequency $\omega_x = 0.8 \pi$ just below the bifurcation,
        all fixed point trajectories perform a single oscillation per period
        roughly orthogonal to the minimum energy path.
        (b)~Decreasing the frequency to $\omega_x = 0.75 \pi$
        leads to the same configuration in principle,
        although with significant changes to the oscillation amplitudes.
        (c)~With $\omega_x = 0.2 \pi$ the situation changes fundamentally.
        The trajectory associated with the central elliptic fixed point
        now follows roughly the reaction coordinate
        approximated by the minimum energy path.
        In doing so, the trajectory always stays near the saddle point.
        Trajectories belonging to the fixed point torus, on the other hand,
        oscillate in direction of the orthogonal mode.
        Furthermore, these trajectories oscillate many times per driving period
        and fan out instead of following the same path back and forth.
        The legend in (b) is shared among all panels.}
    \label{fig:periodic_trajs}
\end{figure}

Figure~\ref{fig:periodic_trajs}a shows the trajectories
associated with the three fixed points in Fig.~\ref{fig:energy_bifurcations}b
just below the bifurcation.
All three trajectories follow very similar paths
roughly orthogonal to the minimum energy path
given by the arctangent in Eq.~\eqref{eq:results/potential}.
The elliptic and hyperbolic fixed points
created in the bifurcation from Fig.~\ref{fig:energy_bifurcations}a
belong to trajectories oscillating in phase.
By contrast, the trajectory associated with the second elliptic fixed point
oscillates in antiphase with a higher amplitude.
This is in accordance with its higher mean energy \ETot\
as shown in Fig.~\ref{fig:periodic_trajs}a.
The slight curvature of the trajectories is caused by the system's nonlinearity.

The nonlinearity is also likely responsible
for the bifurcation at $\omega_x \approx 0.802$:
In the high energy limit $\ETot \to \infty$,
trajectories mostly see an effective potential of the form $V(x, y) = 2 y^2$.
This potential leads to an eigenfrequency of $\omega = 2 \approx 0.637 \pi$
for the orthogonal mode $y$.
For lower energies \ETot, however,
the barrier and the nonlinear coupling
(through the arctangent) need to be taken into account.
As a result, the direction of the orthogonal mode changes to roughly $y - x$
with an increased eigenfrequency.
The system therefore supports a range of frequencies as a function of energy.
Driving the system with a specific frequency $\omega_x$ within this range
selects the corresponding trajectory and associated fixed point.

In Fig.~\ref{fig:energy_bifurcations}a,
two branches show a strong increase of the fixed point energies \ETot\
when the frequency $\omega_x$ is lowered,
quickly rising beyond the limits of numerical feasibility.
This diverging behavior can also be explained by the aforementioned mechanism
since the approximation $V(x, y) = 2 y^2$ is only valid for $\ETot \to \infty$.
The sensitivity of the oscillation amplitude is visible
in the dramatic change in the amplitude of the periodic orbits in
Figs.~\ref{fig:periodic_trajs}a and~\ref{fig:periodic_trajs}b
resulting from the slight reduction
in $\omega_x$ from $0.8 \pi$ to $0.75 \pi$.
The fact that
the oscillation of the smaller amplitude orbit
decreases between the cases in
Figs.~\ref{fig:periodic_trajs}a and Figs.~\ref{fig:periodic_trajs}b
can be attributed to its antiphase oscillation.

Upon lowering the frequency to $\omega_x = 0.2 \pi$
(\cf\ Fig.~\ref{fig:periodic_trajs}c),
only the antiphase fixed point remains.
In contrast to Figs.~\ref{fig:periodic_trajs}a and~\ref{fig:periodic_trajs}b,
its motion now oscillates in phase along the minimum energy path.
In addition, a torus with infinitely many fixed points has emerged.
A typical trajectory associated with one fixed point on the torus
is shown in Fig.~\ref{fig:periodic_trajs}c.
Compared to all the trajectories shown so far,
it oscillates multiple times per period and fans out
instead of remaining on a single periodic path.
These fixed points can be attributed to
a sign change in the torus's winding number.


\subsection{Resonant tori}
\label{sec:results/resonant_tori}

\begin{figure}[tbp]
    \includegraphics[width=\linewidth]{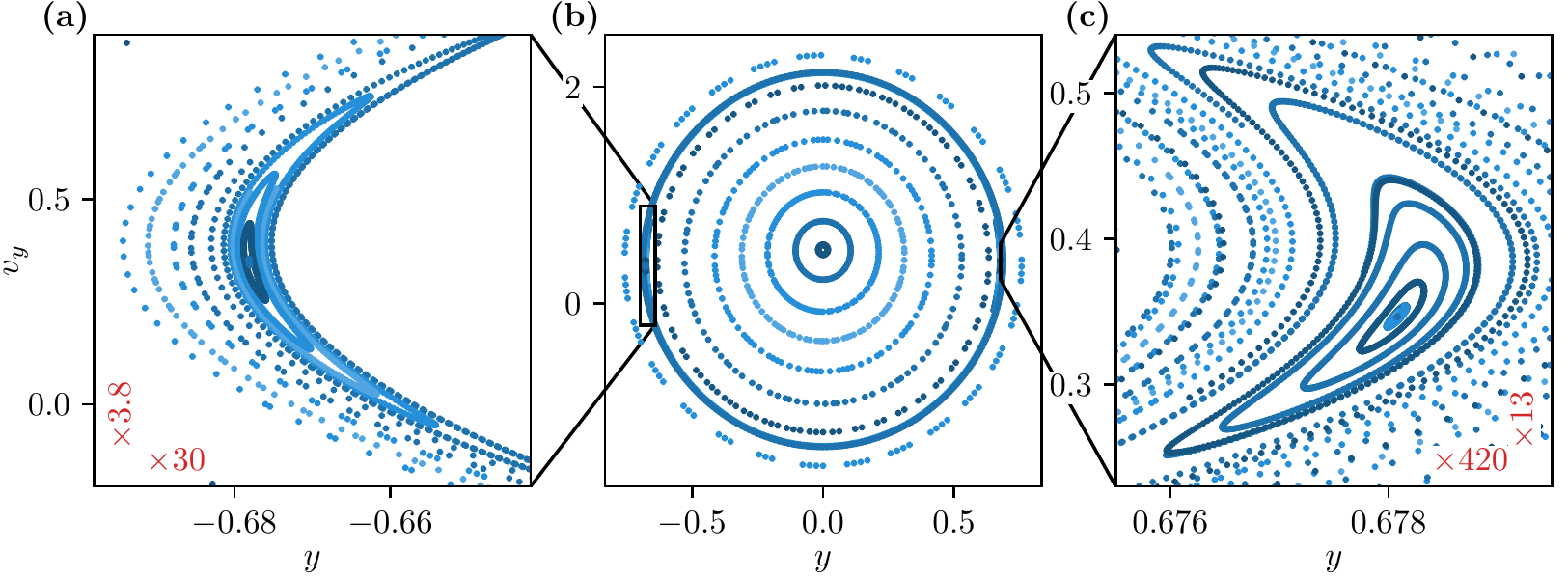}
    \caption{%
        Stroboscopic map of trajectories (\acs{PSOS}) started at time $t_0 = 0$
        for driving frequency $\omega_x = 0.55 \pi$
        and driving amplitude $\hat{x} = 0.4$.
        A color gradient is used to help distinguish neighboring trajectories.
        (b)~From the overview an elliptic fixed point is clearly visible.
        (a)~and~(c) show magnified sections of (b)
        revealing two more elliptic fixed points.
        Different magnifications for the left/right-hand side are used
        since (b) is symmetric under reflection at $y = 0$.
        Labels in the bottom left/right corners indicate
        the approximate $y$ and $v_y$ zoom factors compared to (b).}
    \label{fig:psos_2t_fps}
\end{figure}

Changing the driving frequency to $\omega_x = 0.55 \pi$
results in the occurrence of two elliptic fixed points with period $2 T$.
This can be observed in the \ac{PSOS} by choosing a proper resolution,
as shown in Fig.~\ref{fig:psos_2t_fps}.
Note that the stroboscopic map still records points every $T$ units of time.
The fixed point located at the center of the phase space
(\cf\ Fig.~\ref{fig:psos_2t_fps}b)
is the elliptic fixed point that emerged
from the saddle-node bifurcation in Fig.~\ref{fig:energy_bifurcations}a.

The new fixed points with period $2 T$ can be understood
using the Poincar\'e--Birkhoff
fixed point theorem~\cite{Schuster2005, Wimberger2014}.
It allows the use of the \ac{PSOS} to predict structural changes of resonant tori
when the system is perturbed.
Specifically, a change in the driving frequency can be seen
in Fig.~\ref{fig:psos_2t_fps}
as a perturbation
which causes some of the tori with rational winding numbers to be destroyed.
It results in an even number of fixed points,
alternating between elliptic and hyperbolic ones.
Increasing the perturbation of such a system
can also tear apart the tori in the vicinity of the $2 T$ elliptic fixed points,
forming new fixed points.
Eventually, this leads to self-similar structures.
In principle, one could expect to see
such self-similar structures in Fig.~\ref{fig:psos_2t_fps} when further zooming
the enlarged panels~\ref{fig:psos_2t_fps}a and~\ref{fig:psos_2t_fps}c.
However, we did not continue such zooming here because
resolution is limited by numerical precision.


\section{CONCLUSION AND OUTLOOK}
\label{sec:conclusion}

In this paper,
we have investigated the constrained dynamics on the \ac{NHIM}
in a time-dependent two-dimensional system.
Using Poincar\'e surfaces of section,
we demonstrate the occurrence of structural changes of the dynamics,
\emph{viz.}, bifurcations of periodic
\ac{TS} trajectories when changing the amplitude and frequency of the external driving.
In particular, periodic \ac{TS} trajectories with the same period as the external driving
but significantly different parameters such as mean energy compared to the ordinary
TS trajectory can be created in a saddle-node bifurcation.

The model system investigated in this work,
featuring a periodically driven rank-1-saddle,
is paradigmatic of many chemical reactions in which
the reaction takes place along a reaction path which is
in turn affected by the mode to which it is most
strongly coupled to.
We characterized the dependence of the system's dynamics on the
parameters of the periodic driving at the saddle
\ie, through its frequency and amplitude.
The dynamics of trajectories on the
\ac{NHIM} is unstable because of its proximity to
the rank-1-saddle, but it can nevertheless be obtained
numerically through the use of stabilizing techniques as shown here.
The resulting dynamics was analyzed through stroboscopic maps,
and observed to be regular
for all parameter sets which have been investigated.
At low driving frequencies of the saddle,
a fixed point with a period twice that of the driving was observed.
At higher frequencies, it was also possible to
track the fixed points and observe
the creation and annihilation of pairs of fixed points
in saddle-node-bifurcations.
The behavior of the periodic trajectories was
also projected onto position space leading to the observation
that the saddle frequency
has a significant influence on the oscillation direction of the trajectories.

These results lead to a better understanding
of the reaction dynamics of driven reactions.
Changes in the parameters of the driving
have a huge impact on the system's dynamics
and have been seen here and
recently~\cite{hern20d} to lead to
changes in mean energies and decay rates.
Indeed, it suggests that one could control rate constants within a limited range
by adjusting the driving of the system.

It remains to characterize the behavior
of driven chemical reaction dynamics
as a function of other system parameters,
including the height and width of the saddle
or the various parameters associated with the orthogonal mode.
Changing these parameters could lead to further bifurcations,
paving the way for the possible
observation of a chaotic regime.


\section*{ACKNOWLEDGMENTS}

We thank Tobias Mielich, Robin Bardakcioglu, and Matthias Feldmaier
for fruitful discussions.


\section*{FUNDING}

The German portion of this collaborative work was partially supported by the
Deutsche Forschungsgemeinschaft (DFG) through Grant No.~MA1639/14-1. The US
portion was partially supported by the National Science Foundation (NSF)
through Grant No.~CHE 1700749. This collaboration has also benefited from
support by the European Union's Horizon 2020 Research and Innovation Program
under the Marie Sk\l odowska-Curie Grant Agreement No.~734557.


\bibliography{paper-q29}

\endpaper
\end{document}